\documentclass[aps,prl,twocolumn,showpacs,groupedaddress]{revtex4}
\usepackage{graphicx}
\usepackage{epstopdf}

\begin{document}

%Title of paper
\title{Plasmon-phonon Strongly-Coupled Mode in Epitaxial Graphene}

\author{Yu Liu}
\author{R. F. Willis}
\affiliation{Department of Physics, The Pennsylvania State University, PA 16802 USA}

\date{\today}

\begin{abstract}
We report the dispersion measurements, using angle-resolved reflection electron-energy-loss-spectroscopy (AREELS), on two-dimensional (2D) plasmons in single and multilayer graphene which couple strongly to surface optical phonon (FK phonon) modes of silicon carbide substrate. The coupled modes show discrete dispersion behaviors in the single and bilayer graphene. With increasing graphene layers on SiC(0001), a transition from plasmon-like dispersion to phonon-like dispersion is observed. For plasmon-like modes, the dispersion is strongly damped by electron-hole pair excitations at entering single-particle continuum, while phonon-like mode is undamped. In the region free of coupling, the graphene 2D plasmon exhibits acoustic behavior with linear dispersion with slope and damping determined by the Fermi surface topology.
\end{abstract}

\pacs{71.45.Gm, 73.20.Mf, 73.61.Wp}

\maketitle
The electronic properties of graphene, a two-dimensional (2D) monolayer of carbon atoms connected in a honeycomb lattice, has become the subject of much experimental and theoretical interest since it became practical to produce single sheets of this graphitic material \cite{Geim2007nm}. Due to its high carrier mobility, graphene has the potential to substitute silicon in current silicon-based microelectronics, and the growth of large domain size was already reported on silicon carbide (SiC) wafer \cite{Emtsev2009nm}. The 2D graphene sheet is insolated from insulating SiC substrate, sustaining a strictly 2D electron gas (2DEG) system where the collective oscillation (plasmon) can propagate along the surface. The wavevector-dependent dispersion characterizes the dielectric response of 2DEG, which is important in every aspect of electronic application of graphene. Superior to optical excitation in momentum transfer, angle-resolved electron energy loss spectroscopy (AREELS) provides a unique tool to characterize the dispersion behavior at finite wavevector, where electron-electron correlation effects become evident. The first observation of coupling between surface optical phonon and surface plasmon was on the surface of doped GaAs \cite{Matz1981prl}, the coupling strength was about 20 meV. This dynamic property attracted much attention due to the wide application of GaAs in high electron mobility devices.

We report experimental observation of the coupling between the 2D "sheet plasmon" in graphene and "Fuchs-Kliewer" surface optical phonon (FK phonon) \cite{Fuchs1965pr} in SiC substrate. The "sheet plasmon" is collective oscillation of the $\pi$-valence charge density which propagates within the graphene sheet. This low-energy 2D-plasmon mode is different to the bulk $\pi$ plasmon mode, which is a collective oscillation of the $\pi$-valence electrons in bulk graphite \cite{Pichler1998prl} and coexisted in our spectrum measurements. The graphene $\pi$-plasmon mode is a strictly 2D mode whose energy decreases continuously to zero in the longwavelength (classical) limit $q\to0$. FK phonon arising from displacement of lattice ions is normally observed at polar semiconductor surfaces, such as GaAs and GaP \cite{Dubois1982prb}.

Our previous measurements \cite{Liu2008prb} showed a discrepancy at small wavevector ($q<$0.05\AA$^{-1}$); the graphene sheet plasmon energy deviating strongly from acoustic behavior. Similar measurements on the base 6H-SiC(0001) substrate showed the presence of a large-amplitude FK phonon mode in this same energy-wavevector range \cite{Dubois1982prb, Nienhaus1995ss}. The suspicion was that the two modes might be strongly coupled.

In an effort to resolve any coupling, we tuned the smallest possible incident electron energy which was high enough to excite plasmon and had a strong intensity for off-specular measurement, so as to increase the momentum resolution of our measurements. We observed an unusually strong coupling about 130 meV between the sheet plasmon in the graphene and the FK phonon in the SiC substrate. We report these measurements here. Also, we compared the behavior with that of increasing number of graphene sheets, 1 to 5 monolayers (ML), grown on the same SiC(0001) substrate. The observed strong coupling carries over into  each layer. However, subtle changes in the damping and dispersion are observed, due to a more complex Fermi surface, the topology and carrier density changing from strictly 2D graphene to 3D graphite.

\begin{figure}
\centering
\includegraphics[width=3.2in]{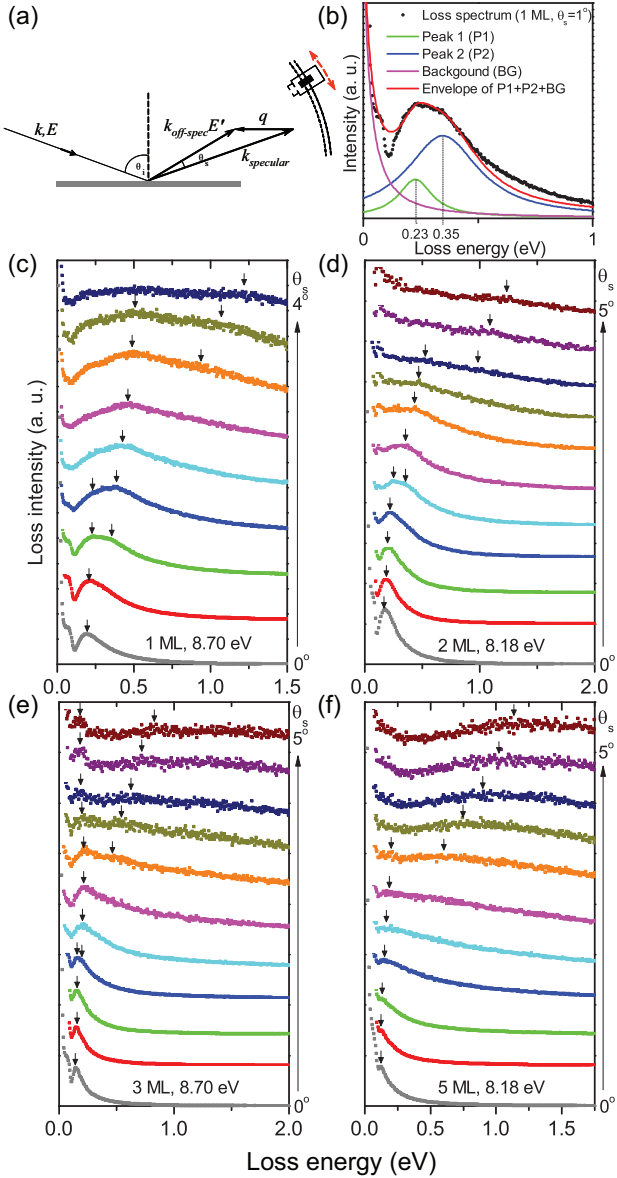}
\caption{(a) scattering geometry of AREELS experiments, $\theta_\mathrm{i}$ = 60$^\circ$. (b) the extraction of loss peaks in spectrum taken at $\theta_\mathrm{s}$=1$^\circ$ on 1 ML graphene, which is comprised of two Lorentz peaks and background. Loss spectra of (c) 1.2 ML, (d) 2.3 ML, (e) 3.3 ML and (f) 4.7 ML graphene dispersing with increasing scattering angle $\theta_\mathrm{S}$. Black arrows indicate the position of loss peaks after the subtraction of background and the deconvolution of peak envelope composing of two peaks} \label{fig:spectra}
\end{figure}

Graphene layers were prepared on a 6H-SiC(0001) crystalline wafer surface (0.1 $\Omega$ cm resistivity) by solid state graphitization \cite{Forbeaux1998prb}. The sheet structure has been well characterized and floating on a carbon-rich $(6\sqrt 3 \times 6 \sqrt 3)R30^{\circ}$ interfacial superstructure \cite{Bostwick2007np, Emtsev2007msf}. AREELS measurements were performed at room temperature and in ultrahigh vacuum ($2\times10^{-10}$ Torr base pressure), on graphene layers of four different thicknesses and on a bare hydrogen-etched 6H-SiC(0001) surface. A medium energy resolution of 10 meV was set to achieve strong signal, and the momentum resolution was better than 0.01 \AA$^{-1}$. \footnote{The instrument was a commercial LK2000-14-R spectrometer (LK Technologies) with slits of 0.13 mm of two primary monochromators. The angular acceptance $\alpha$ is 0.15$^\circ$, the estimated momentum resolution $\Delta q$ is given by $\sqrt{2mE_\mathrm{i}/\hbar}(\mathrm{cos}\theta_\mathrm{i}+\sqrt{1-E_{\mathrm{loss}}/E_\mathrm{i}}\mathrm{cos}\theta_s)\alpha$ (Ref. \cite{Rocca1995ssr}). Typical setting for our spectrometer: the impact energy $E_\mathrm{i}$=8 eV, $\theta_\mathrm{i}$=60$^\circ$, and loss energy $E_\mathrm{loss} \sim$ 200 meV around specular direction, the $\Delta q$ = 0.0038 \AA$^{-1}$; at 5$^\circ$ away from specular direction ($\theta_\mathrm{s}$=5$^\circ$), $\Delta q$ = 0.0040 \AA$^{-1}$.}, which is small enough to resolve the concentrated peak profile of dipole scattering \cite{Liu2008prb}. The wavevector $q$ of each measurement at a particular scattering angle $\theta_\mathrm{s}$ is calculated by following equation reflecting the conservation of energy and momentum in plane,
\begin{equation}\label{eqn:momentum}
q=\frac{\sqrt{2m_\mathrm{e}E}}{\hbar}[\mathrm{sin}\theta_\mathrm{i}-\sqrt{1-E_\mathrm{loss}/E}\;\mathrm{sin}(\theta_\mathrm{i}-\theta_\mathrm{s})].
\end{equation}
where the loss energy $E_\mathrm{loss} = E - E'$  and other definitions of variables are illustrated in Fig. \ref{fig:spectra}(a). The step between each measurement is determined by the impact energy $E$ and the scattering angle $\theta_\mathrm{s}$ which has a minimum step of 0.5$^\circ$. A smaller step of wavevector can be achieved with smaller impact energy $E$, however the strength of spectrum signals also varies with impact energy $E$ badly, so a considerable effort on tuning is needed to find the optimum setting to not only distinguish fine structures in $q$ space, but also to yield enough signal strength at large off-specular scattering. In our work, the optimum setting for our spectrometer is $E \sim8$ eV.

Fig. \ref{fig:spectra} (c)-(f) show the energy loss peaks in the AREELS spectrum of low energy electrons back-scattered from graphene of various thicknesses on SiC(0001). The black arrows in Fig. \ref{fig:spectra}(c)-(f) indicate the peak positions in loss spectra, which are the superpositions of multiple loss structures and the background, shown in Fig. \ref{fig:spectra}(b). Generally the loss peaks disperse with increasing scattering angle $\theta_\mathrm{s}$, corresponding to increasing wavevector parallel to the surface. The surprising spectroscopic structure is the splittings of peaks at certain energy, which is not disclosed in measurements with larger step at higher impact energy. The substrate of our samples, SiC, is a polarized ionic crystal with large band gap, and has attracted much attention in last decades, including researches on its surface excitation of various surface structures \cite{Harris1995b}. In order to clarify the origins of this unexpected splitting of graphene dispersion, we took measurements on a bare 6H-SiC(0001) surface, which is different to 1 ML graphene only in top two layers, i.e. $6\sqrt{3}$ buffer layer and top graphene layer, \cite{Emtsev2007msf} so that, the polarization contributed from SiC substrate can be identified separately and used to analyzes measurements on graphene.

\begin{figure}
\centering
\includegraphics[width=2.8in]{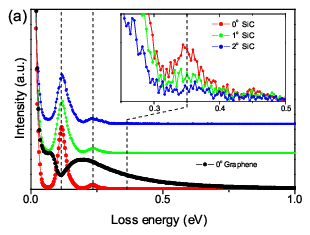}
\includegraphics[width=2.0in]{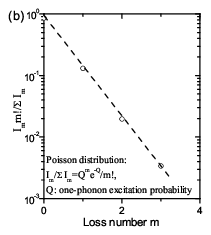}
\caption{(a) loss spectra on SiC surface at various angles and normalized specular spectrum of 1 ML graphene (black). The loss spectra on SiC show multi-phonon loss peaks at 117 meV, 236 meV, and 351 meV (magnified in insert). (b) the Poisson distribution of intensities of the specular spectrum} \label{fig:SiC}
\end{figure}
Fig. \ref{fig:SiC}(a) shows AREELS spectrum taken on a bare 6H-SiC(0001) surface before growing the graphene layer. Each spectrum of SiC show equally spaced peaks and the probability of multi-phonon excitation follows the Poisson distribution, shown in Fig. \ref{fig:SiC}(b), which is the characteristic of multi-phonon process \cite{Ibach1970prl}. The calculated FK phonon frequency $\omega_\mathrm{FK}$ is 116.7 meV \cite{Dubois1982prb, Nienhaus1995ss}, in good agreement with experimental measurement of one-phonon excitation, 117 $\pm$ 10 meV and its multiples in the process of multi-phonon excitations. Under our experimental resolution ($\sim$ 10 meV), those FK-phonon peaks show little dispersion (straight dashed line in Fig. \ref{fig:SiC}(a)), in contrast to the large shifting of peaks on graphene layers (energy shifting $\Delta \sim$ 1 eV). As a comparison, Fig. \ref{fig:SiC}(a) also shows a normalized spectrum (black) taken on a single layer graphene in specular direction ($\theta_\mathrm{s}$ = 0$^\circ$) at the same impact energy, where a drastic change of loss structure close to the elastic peak is due to the total reformation of atomic structure in the SiC(0001) surface.

\begin{figure}
\centering
\includegraphics[width=2.8in]{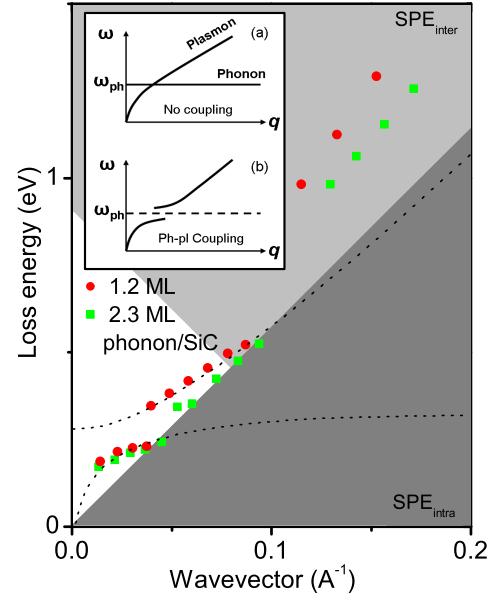}
\caption{The energy of Loss peaks plotted as a function of wavevector q calculated by Eqn. \ref{eqn:momentum}. The impact energy is around 8 eV. The dashed curves are guidelines for two dispersion branches. Two shaded areas are single-particle excitations (SPE) continuum due to intra-band and inter-band transition. Insert: schematic of the coupling at which plasmons disperse across FK phonon modes.}\label{fig:dispersion1}
\end{figure}

By using low-energy reflection electron energy loss spectroscopy with its shallow probing depth in the surface region, the collected loss structure in spectrum reflects a long range dipole field decaying quickly in the bulk side. The excitation of FK phonons on SiC surface is actually the oscillation of induced field due to the displacement of positive and negative ions, which bears a similar dielectric response properties to dipole field of 2D plasmons induced by collective oscillation of electrons. This dielectric similarity enable the coupling in the region where the energy of FK phonons is comparative with that of 2D plasmon. But these two modes have very different origins and causes different screening behavior upon entering the single-particle excitation continuum, Fig. \ref{fig:dispersion1}.

Fig. \ref{fig:dispersion1} shows the dispersion behavior of the loss peaks observed in the AREELS spectra taken on graphene of various thicknesses. The distribution of single-particle excitations (SPE) due to intra- and inter-band transition of 2DEG is also plotted with Fermi energy $E_\mathrm{F}$ = 0.45 eV, and estimated Fermi wavevector $k_\mathrm{F}$=0.079 \AA$^{-1}$ at electron density $n=2 \times 10^{13}$cm$^{-2}$ \cite{Bostwick2007np}. For 1.2 ML graphene (red dot in Fig. \ref{fig:dispersion1}), instead of smoothly increasing, the dispersion loses its continuity and jumps at $q \sim$ 0.05 \AA$^{-1}$ where crossovers occurs between plasmons and FK phonons ($\omega_\mathrm{FK} >$117 meV) . The insert in Fig. \ref{fig:dispersion1} shows a schematic change of dispersion curve where two modes couple with each other. Like the damping of plasmon dispersion of classical 2DEG, the cutoffs at entering SPE continuum are also observed in both separated branches, indicating that the coupled mode is dominated by plasmon feature, i.e. plasmon-like. Comparing with 1.2 ML graphene, 2.3 ML graphene (green square) exhibits similarly in splitting loss spectra and discrete dispersion. The slight difference comes from the second branch ($0.05 <q < 0.1$\AA$^{-1}$ ), where the trend of data points is not as predictable as that of 1.2 ML, and can not be analyzed effectively if no more points are obtained with higher angular resolution of AREELS. The possible reason for this deviation is that electrons participating collective oscillation, which are also those electrons filled in Fermi level crossing $\pi$ bands, are affected by a splitting of $\pi$ bands due to the formation of the second graphene layer \cite{Ohta2007prl}.

\begin{figure}
\centering
\includegraphics[width=2.8in]{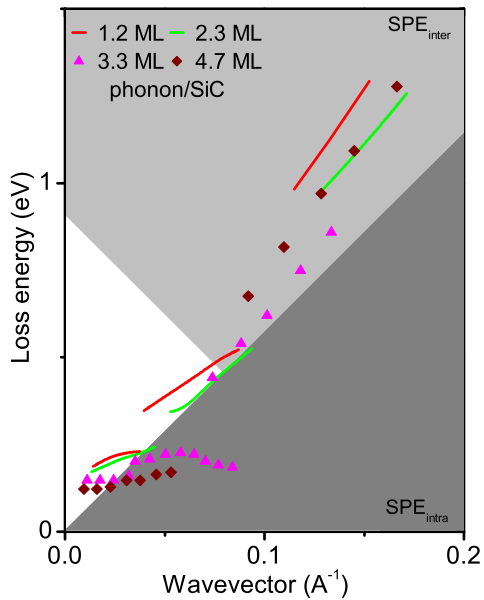}
\caption{Dispersion behaviors of 3 ML (pink triangles) and 5 ML (brown diamonds) graphene on SiC, in comparison with that of 1 ML (red line) and 2 ML (green line).}\label{fig:dispersion2}
\end{figure}
With more graphene layers grown on SiC, the dispersion exhibits dramatic change. In Fig. \ref{fig:dispersion2}, the dispersion of 3.3 ML graphene(pink triangles) and 4.7 ML (brown diamond) graphene are plotted with that of single (red line) and bilayer (green line) graphene to show the difference. Firstly, the energy and energy shifting of the coupled mode in 3.3 ML and 4.7 ML graphene are much smaller than that of single and bilayer graphene. The dispersions of 3.3 ML and 4.7 ML graphene are contained between one-phonon and two-phonon process and much flatter. Secondly, the dispersion continues into SPE, in contrast to the cutoffs observed in single and bilayer graphene. This indicates a decreasing weight of plasmon feature in the coupled mode, consequently the dispersion turns to phonon-like which appears a much flatter dispersion and free of damping from electron-hole pair excitation in SPE. Accumulated with added graphene layer, the interaction between graphene layers increases, the characteristic of plasmon-phonon mode transits from plasmon-like to phonon-like. Basically this transition is due to the weakening of two-dimensional confinement of electron gas.

Besides the discrete dispersion in low-$q$ ($q < 0.1$ \AA$^{-1}$ ) region due to plasmon-phonon coupling, another interesting observation is the linear dispersion in large-$q$ ($q > 0.1$\AA$^{-1}$). For an ideal 2DES, its plasmon frequency $\omega_\mathrm{2D}$ follows $\sqrt q$ dispersion in the local-field approximation in the longwavelength limit \cite{March1995ap}. Stern \cite{Stern1967prl} derived a higher-order correction including the effect of non-local field in a random-phase-approximation:
\begin{eqnarray}\label{equ:Stern}
\omega_\mathrm{2D}=[\frac{4 \pi ne^2}{m^*(1+\epsilon_\mathrm{s})}|q|+\frac{3}{4}v_\mathrm{F}^2q^2+...]^{\frac{1}{2}}
\end{eqnarray}
where $n$, $m^*$, and $\epsilon_\mathrm{s}$ are the areal electron density, effective mass and static dielectric constant of the medium, respectively. The first-order correction expresses in terms of the Fermi velocity $v_\mathrm{F}$. At finite $q$, this correction term is dominant, plasmon disperse linearly. However, we note the recent discovery on metal surface, a so-called acoustic surface plasmon also shows a linear dispersion with \cite{Diaconescu2007n}:
\begin{eqnarray}\label{equ:Stern}
\omega_\mathrm{2D}= \alpha v_\mathrm{F} q,
\end{eqnarray}
where coefficient $\alpha$ depends on the electron density of substrate and distance between 2D sheet and substrate. This linearity of plasmon is due to out-of-phase oscillation between electrons in 2D sheet and in substrate. In our case, the epitaxial graphene bears a similar structure, i.e. strictly 2D graphene sheet floating on a doped semiconductor substrate. We also notice the change of slope of dispersions for different graphene layers in large-$q$ region, corresponding to the change of group velocity of plasmon propagation, $v_\mathrm{F}$. This is due to the scattering from degenerate $\pi$ bands induced by extra graphene layers. Furthermore, the change of slope is not monotonously decreasing, as the slope of 4.7 ML is larger than 3.3 ML in Fig. \ref{fig:dispersion2}, indicative of the existence of in-phase and out-of-phase polarization between layers which can be reflected in the oscillating charge distribution of multiple graphene layers \cite{Guinea2007prb}. The dispersion and damping in this region ($q>0.1$\AA$^{-1}$) also reflects the changing topology of the Fermi surface, evolving from  2D graphene to 3D graphite.

In conclusion, we have observed an unusually strong coupling ($\sim$130 meV) between the dipolar electric fields generated by oscillations of the sheet charge of the $\pi$ electrons in graphene and the ionic charges in the substrate SiC. The sheet-charge density $n \sim 10^{13}$cm$^{-2}$ is a result of charge transfer out of surface states on the SiC into empty $\pi^*$ states in the graphene sheet. Fuchs-Kliewer surface optical phonon modes in the SiC cause this charge transfer to oscillate. We believe this to be the origin of the unusually strong splitting observed. An usual feature, is that this coupling extends to graphene multilayers i.e. the multilayers are in noway screened by the first-layer, despite its very metallic character. This suggests charge transfer/scattering between layers. Further evidence is provided by changes in the plasmon dispersion of higher wavevectors $q>0.1$\AA$^{-1}$. However, the single-particle damping of the coupled mode does show different behavior, the dispersion continuing into the single-particle continuum in the case of the multilayers. This is indicative of subtle changes in momentum space, i.e. change in Fermi surface topology evolving from 2D graphene to 3D graphite, and changes in the screening properties.\\

We would like to thank Dr. Thomas Seyller for preparing graphene samples. We also thank the Petroleum Research Fund of the American Physical Society for funding this research, Grant No. 43557-AC5.


\begin{thebibliography}{19}
\expandafter\ifx\csname natexlab\endcsname\relax\def\natexlab#1{#1}\fi
\expandafter\ifx\csname bibnamefont\endcsname\relax
  \def\bibnamefont#1{#1}\fi
\expandafter\ifx\csname bibfnamefont\endcsname\relax
  \def\bibfnamefont#1{#1}\fi
\expandafter\ifx\csname citenamefont\endcsname\relax
  \def\citenamefont#1{#1}\fi
\expandafter\ifx\csname url\endcsname\relax
  \def\url#1{\texttt{#1}}\fi
\expandafter\ifx\csname urlprefix\endcsname\relax\def\urlprefix{URL }\fi
\providecommand{\bibinfo}[2]{#2}
\providecommand{\eprint}[2][]{\url{#2}}

\bibitem[{\citenamefont{Geim and Novoselov}(2007)}]{Geim2007nm}
\bibinfo{author}{\bibfnamefont{A.~K.} \bibnamefont{Geim}} \bibnamefont{and}
  \bibinfo{author}{\bibfnamefont{K.~S.} \bibnamefont{Novoselov}},
  \bibinfo{journal}{Nature Materials} \textbf{\bibinfo{volume}{6}},
  \bibinfo{pages}{183} (\bibinfo{year}{2007}).

\bibitem[{\citenamefont{Emtsev et~al.}(2009)\citenamefont{Emtsev, Bostwick,
  Horn, Jobst, Kellogg, Ley, McChesney, Ohta, Reshanov, Rohrl
  et~al.}}]{Emtsev2009nm}
\bibinfo{author}{\bibfnamefont{K.~V.} \bibnamefont{Emtsev}},
  \bibinfo{author}{\bibfnamefont{A.}~\bibnamefont{Bostwick}},
  \bibinfo{author}{\bibfnamefont{K.}~\bibnamefont{Horn}},
  \bibinfo{author}{\bibfnamefont{J.}~\bibnamefont{Jobst}},
  \bibinfo{author}{\bibfnamefont{G.~L.} \bibnamefont{Kellogg}},
  \bibinfo{author}{\bibfnamefont{L.}~\bibnamefont{Ley}},
  \bibinfo{author}{\bibfnamefont{J.~L.} \bibnamefont{McChesney}},
  \bibinfo{author}{\bibfnamefont{T.}~\bibnamefont{Ohta}},
  \bibinfo{author}{\bibfnamefont{S.~A.} \bibnamefont{Reshanov}},
  \bibinfo{author}{\bibfnamefont{J.}~\bibnamefont{Rohrl}},
  \bibnamefont{et~al.}, \bibinfo{journal}{Nat Mater}
  \textbf{\bibinfo{volume}{8}}, \bibinfo{pages}{203} (\bibinfo{year}{2009}).

\bibitem[{\citenamefont{Matz and L\"uth}(1981)}]{Matz1981prl}
\bibinfo{author}{\bibfnamefont{R.}~\bibnamefont{Matz}} \bibnamefont{and}
  \bibinfo{author}{\bibfnamefont{H.}~\bibnamefont{L\"uth}},
  \bibinfo{journal}{Phys. Rev. Lett.} \textbf{\bibinfo{volume}{46}},
  \bibinfo{pages}{500} (\bibinfo{year}{1981}).

\bibitem[{\citenamefont{Fuchs and Kliewer}(1965)}]{Fuchs1965pr}
\bibinfo{author}{\bibfnamefont{R.}~\bibnamefont{Fuchs}} \bibnamefont{and}
  \bibinfo{author}{\bibfnamefont{K.~L.} \bibnamefont{Kliewer}},
  \bibinfo{journal}{Phys. Rev.} \textbf{\bibinfo{volume}{140}},
  \bibinfo{pages}{A2076} (\bibinfo{year}{1965}).

\bibitem[{\citenamefont{Pichler et~al.}(1998)\citenamefont{Pichler, Knupfer,
  Golden, Fink, Rinzler, and Smalley}}]{Pichler1998prl}
\bibinfo{author}{\bibfnamefont{T.}~\bibnamefont{Pichler}},
  \bibinfo{author}{\bibfnamefont{M.}~\bibnamefont{Knupfer}},
  \bibinfo{author}{\bibfnamefont{M.~S.} \bibnamefont{Golden}},
  \bibinfo{author}{\bibfnamefont{J.}~\bibnamefont{Fink}},
  \bibinfo{author}{\bibfnamefont{A.}~\bibnamefont{Rinzler}}, \bibnamefont{and}
  \bibinfo{author}{\bibfnamefont{R.~E.} \bibnamefont{Smalley}},
  \bibinfo{journal}{Phys. Rev. Lett.} \textbf{\bibinfo{volume}{80}},
  \bibinfo{pages}{4729} (\bibinfo{year}{1998}).

\bibitem[{\citenamefont{Dubois and Schwartz}(1982)}]{Dubois1982prb}
\bibinfo{author}{\bibfnamefont{L.~H.} \bibnamefont{Dubois}} \bibnamefont{and}
  \bibinfo{author}{\bibfnamefont{G.~P.} \bibnamefont{Schwartz}},
  \bibinfo{journal}{Physical Review B} \textbf{\bibinfo{volume}{26}},
  \bibinfo{pages}{794} (\bibinfo{year}{1982}).

\bibitem[{\citenamefont{Liu et~al.}(2008)\citenamefont{Liu, Willis, Emtsev, and
  Seyller}}]{Liu2008prb}
\bibinfo{author}{\bibfnamefont{Y.}~\bibnamefont{Liu}},
  \bibinfo{author}{\bibfnamefont{R.~F.} \bibnamefont{Willis}},
  \bibinfo{author}{\bibfnamefont{K.~V.} \bibnamefont{Emtsev}},
  \bibnamefont{and} \bibinfo{author}{\bibfnamefont{T.}~\bibnamefont{Seyller}},
  \bibinfo{journal}{Phys. Rev. B Rapid Comm.} \textbf{\bibinfo{volume}{78}},
  \bibinfo{pages}{201403} (\bibinfo{year}{2008}).

\bibitem[{\citenamefont{Nienhaus et~al.}(1995)\citenamefont{Nienhaus, Kampen,
  and M?ch}}]{Nienhaus1995ss}
\bibinfo{author}{\bibfnamefont{H.}~\bibnamefont{Nienhaus}},
  \bibinfo{author}{\bibfnamefont{T.~U.} \bibnamefont{Kampen}},
  \bibnamefont{and} \bibinfo{author}{\bibfnamefont{W.}~\bibnamefont{M?ch}},
  \bibinfo{journal}{Surface Science} \textbf{\bibinfo{volume}{324}},
  \bibinfo{pages}{L328} (\bibinfo{year}{1995}).

\bibitem[{\citenamefont{Forbeaux et~al.}(1998)\citenamefont{Forbeaux, Themlin,
  and Debever}}]{Forbeaux1998prb}
\bibinfo{author}{\bibfnamefont{I.}~\bibnamefont{Forbeaux}},
  \bibinfo{author}{\bibfnamefont{J.-M.} \bibnamefont{Themlin}},
  \bibnamefont{and} \bibinfo{author}{\bibfnamefont{J.-M.}
  \bibnamefont{Debever}}, \bibinfo{journal}{Phys. Rev. B}
  \textbf{\bibinfo{volume}{58}}, \bibinfo{pages}{16396} (\bibinfo{year}{1998}).

\bibitem[{\citenamefont{Bostwick et~al.}(2007)\citenamefont{Bostwick, Ohta,
  Seyller, Horn, and Rotenberg}}]{Bostwick2007np}
\bibinfo{author}{\bibfnamefont{A.}~\bibnamefont{Bostwick}},
  \bibinfo{author}{\bibfnamefont{T.}~\bibnamefont{Ohta}},
  \bibinfo{author}{\bibfnamefont{T.}~\bibnamefont{Seyller}},
  \bibinfo{author}{\bibfnamefont{K.}~\bibnamefont{Horn}}, \bibnamefont{and}
  \bibinfo{author}{\bibfnamefont{E.}~\bibnamefont{Rotenberg}},
  \bibinfo{journal}{Nature Physics} \textbf{\bibinfo{volume}{3}},
  \bibinfo{pages}{36} (\bibinfo{year}{2007}).

\bibitem[{\citenamefont{Emtsev et~al.}(2007)\citenamefont{Emtsev, Seyller,
  Speck, Ley, Stojanov, Riley, and Leckey}}]{Emtsev2007msf}
\bibinfo{author}{\bibfnamefont{K.~V.} \bibnamefont{Emtsev}},
  \bibinfo{author}{\bibfnamefont{T.}~\bibnamefont{Seyller}},
  \bibinfo{author}{\bibfnamefont{F.}~\bibnamefont{Speck}},
  \bibinfo{author}{\bibfnamefont{L.}~\bibnamefont{Ley}},
  \bibinfo{author}{\bibfnamefont{P.}~\bibnamefont{Stojanov}},
  \bibinfo{author}{\bibfnamefont{J.}~\bibnamefont{Riley}}, \bibnamefont{and}
  \bibinfo{author}{\bibfnamefont{R.}~\bibnamefont{Leckey}},
  \bibinfo{journal}{Mater. Sci. Forum} \textbf{\bibinfo{volume}{556}},
  \bibinfo{pages}{525} (\bibinfo{year}{2007}).

\bibitem[{\citenamefont{Harris}(1995)}]{Harris1995b}
\bibinfo{author}{\bibfnamefont{G.~L.} \bibnamefont{Harris}},
  \emph{\bibinfo{title}{Properties of silicon carbide}}
  (\bibinfo{publisher}{INSPEC, Institution of Electrical Engineers},
  \bibinfo{address}{London}, \bibinfo{year}{1995}).

\bibitem[{\citenamefont{Ibach}(1970)}]{Ibach1970prl}
\bibinfo{author}{\bibfnamefont{H.}~\bibnamefont{Ibach}},
  \bibinfo{journal}{Phys. Rev. Lett.} \textbf{\bibinfo{volume}{24}},
  \bibinfo{pages}{1416} (\bibinfo{year}{1970}).

\bibitem[{\citenamefont{Ohta et~al.}(2007)\citenamefont{Ohta, Bostwick,
  McChesney, Seyller, Horn, and Rotenberg}}]{Ohta2007prl}
\bibinfo{author}{\bibfnamefont{T.}~\bibnamefont{Ohta}},
  \bibinfo{author}{\bibfnamefont{A.}~\bibnamefont{Bostwick}},
  \bibinfo{author}{\bibfnamefont{J.~L.} \bibnamefont{McChesney}},
  \bibinfo{author}{\bibfnamefont{T.}~\bibnamefont{Seyller}},
  \bibinfo{author}{\bibfnamefont{K.}~\bibnamefont{Horn}}, \bibnamefont{and}
  \bibinfo{author}{\bibfnamefont{E.}~\bibnamefont{Rotenberg}},
  \bibinfo{journal}{Phys. Rev. Lett.} \textbf{\bibinfo{volume}{98}},
  \bibinfo{pages}{206802} (\bibinfo{year}{2007}).

\bibitem[{\citenamefont{March and Tosi}(1995)}]{March1995ap}
\bibinfo{author}{\bibfnamefont{N.~H.} \bibnamefont{March}} \bibnamefont{and}
  \bibinfo{author}{\bibfnamefont{M.~P.} \bibnamefont{Tosi}},
  \bibinfo{journal}{Advances in Physics} \textbf{\bibinfo{volume}{44}},
  \bibinfo{pages}{299} (\bibinfo{year}{1995}).

\bibitem[{\citenamefont{Stern}(1967)}]{Stern1967prl}
\bibinfo{author}{\bibfnamefont{F.}~\bibnamefont{Stern}},
  \bibinfo{journal}{Phys. Rev. Lett.} \textbf{\bibinfo{volume}{18}},
  \bibinfo{pages}{546} (\bibinfo{year}{1967}).

\bibitem[{\citenamefont{Diaconescu et~al.}(2007)\citenamefont{Diaconescu, Pohl,
  Vattuone, Savio, Hofmann, Silkin, Pitarke, Chulkov, Echenique, Farias
  et~al.}}]{Diaconescu2007n}
\bibinfo{author}{\bibfnamefont{B.}~\bibnamefont{Diaconescu}},
  \bibinfo{author}{\bibfnamefont{K.}~\bibnamefont{Pohl}},
  \bibinfo{author}{\bibfnamefont{L.}~\bibnamefont{Vattuone}},
  \bibinfo{author}{\bibfnamefont{L.}~\bibnamefont{Savio}},
  \bibinfo{author}{\bibfnamefont{P.}~\bibnamefont{Hofmann}},
  \bibinfo{author}{\bibfnamefont{V.~M.} \bibnamefont{Silkin}},
  \bibinfo{author}{\bibfnamefont{J.~M.} \bibnamefont{Pitarke}},
  \bibinfo{author}{\bibfnamefont{E.~V.} \bibnamefont{Chulkov}},
  \bibinfo{author}{\bibfnamefont{P.~M.} \bibnamefont{Echenique}},
  \bibinfo{author}{\bibfnamefont{D.}~\bibnamefont{Farias}},
  \bibnamefont{et~al.}, \bibinfo{journal}{Nature}
  \textbf{\bibinfo{volume}{448}}, \bibinfo{pages}{57} (\bibinfo{year}{2007}).

\bibitem[{\citenamefont{Guinea}(2007)}]{Guinea2007prb}
\bibinfo{author}{\bibfnamefont{F.}~\bibnamefont{Guinea}},
  \bibinfo{journal}{Phys. Rev. B} \textbf{\bibinfo{volume}{75}},
  \bibinfo{pages}{235433} (\bibinfo{year}{2007}).

\bibitem[{\citenamefont{Rocca}(1995)}]{Rocca1995ssr}
\bibinfo{author}{\bibfnamefont{M.}~\bibnamefont{Rocca}},
  \bibinfo{journal}{Surf. Sci. Rep.} \textbf{\bibinfo{volume}{22}},
  \bibinfo{pages}{1} (\bibinfo{year}{1995}).

\end{thebibliography}
\end{document}